# Fight Hardware with Hardware: System-wide Detection and Mitigation of Side-Channel Attacks using Performance Counters

STEFANO CARNÀ, Sapienza, University of Rome, Italy
SERENA FERRACCI, Sapienza, University of Rome, Italy
FRANCESCO QUAGLIA*, University of Rome "Tor Vergata", Italy
ALESSANDRO PELLEGRINI[†‡], University of Rome "Tor Vergata", Italy

We present a kernel-level infrastructure that allows system-wide detection of malicious applications attempting to exploit cache-based side-channel attacks to break the process confinement enforced by standard operating systems. This infrastructure relies on hardware performance counters to collect information at runtime from all applications running on the machine. High-level detection metrics are derived from these measurements to maximize the likelihood of promptly detecting a malicious application. Our experimental assessment shows that we can catch a large family of side-channel attacks with a significantly reduced overhead. We also discuss countermeasures that can be enacted once a process is suspected of carrying out a side-channel attack to increase the overall tradeoff between the system's security level and the delivered performance under non-suspected process executions.

CCS Concepts: • **Security and privacy** → **Hardware-based security protocols**; **Operating systems security**; *Software security engineering*.

Additional Key Words and Phrases: side-channel attack, covert channel, detection, mitigation, hardware performance counters, operating systems.

## 1 INTRODUCTION

Historically, the memory hierarchy has been introduced and equipped with multiple caches to hide the memory's high latency and overcome the performance gap between processors and memory. This component is fundamental performance-wise and has been integrated into all modern processing units. Nevertheless, attackers have repeatedly abused it to extract information from the operating system (OS) kernel or victim processes, circumventing the process confinement enforced by standard modern operating systems.

These kinds of attacks are commonly and generically referred to as *cache-based side-channel attacks*, and many different techniques have been proposed in the literature [16, 22, 33, 36, 37, 59]. They rely on observing non-functional properties of the cache architecture (e.g., timing the execution of cache-accessing operations)

---

*Also with National Inter-University Consortium for Telecommunications (CNIT).
†Also with National Inter-University Consortium for Telecommunications (CNIT).
‡Corresponding author.

Authors' addresses: Stefano Carnà, carna@diag.uniroma1.it, Sapienza, University of Rome, Rome, Italy; Serena Ferracci, ferracci@diag.uniroma1.it, Sapienza, University of Rome, Rome, Italy; Francesco Quaglia, francesco.quaglia@uniroma2.it, University of Rome "Tor Vergata", Rome, Italy; Alessandro Pellegrini, a.pellegrini@ing.uniroma2.it, University of Rome "Tor Vergata", Rome, Italy.

to infer information on a victim process or to leak data. These attacks have also been beneficial to extract data when side effects on the cache architecture are generated by exploiting transient execution CPU vulnerabilities. To exploit these vulnerabilities, the attacker relies on some speculative execution facility from the underlying computer architecture to load into the cache architecture some data from the underlying OS kernel or from a different userspace application whose access is prevented by traditional security mechanisms, such as paging isolation.

The applications of side-channel attacks are vast. For example, they have been used to extract secret keys from cryptographic algorithms including AES [28, 44, 51] or El-Gamal [37], to steal information from the underlying operating system [33, 36], to bypass Kernel Address Space Layout Randomization (KASLR) [38], or to extract cross-VM information [8, 10, 52, 58].

The idea of using a cache-based side channel to trace the execution of a program or to leak information is not new, with the first proposals dating back to two decades ago [2, 7, 44, 62]. Despite this, the fundamental mechanism has been preserved over time, i.e., observing traces left in the caching subsystem of the computer architecture to extract information. In this article, we concentrate on detecting an application mounting a side channel to indicate that an information extraction might be taking place. As we will show experimentally, this focus allows us to detect also transient execution attacks if they rely on side channels to extract leaked information.

To reduce the impact of attacks that exploit cache side channels, especially those based on CPU transient execution—namely to reduce the amount of information that an attacker can leak —several mitigation strategies have been proposed, both at the software and at the hardware level. The most notable ones are Kernel Page Table Isolation (KPTI) [18], retpolines [1], `swapgs` fences, or PCID. Notably, some of these patches (e.g., KPTI) induce a non-negligible performance drop under specific workloads, which has been estimated as high as 30% [21]. This overhead could be deemed too high in specific scenarios—examples are virtualized environments supporting 5G communications [8] and high-performance computing-oriented setups. In these scenarios, a good tradeoff could be to selectively enable security patches (either the existing ones or newly devised ones) at runtime only when software suspected to try to exploit cache side channels is detected.

To detect an application mounting a side channel with a reduced performance impact, we propose to leverage hardware capabilities offered by off-the-shelf CPUs. Indeed, modern CPUs are equipped with hardware units meant to profile the performance or (to some extent) the behavior of applications—they have been introduced long ago in processor families like the Intel Pentium or the AMD Athlon. These units are generally referred to as Performance Monitoring Units (PMUs). PMUs, at their simplest, are composed of programmable Performance Monitor Counters (PMCs), also referred to as Hardware Performance Counters (HPCs). HPCs can be used to keep track of the number of micro-architectural events which occur in the system, such as the number of load operations retired by CPU cores or the number of cache misses at a particular cache level.

Given the tight connection with the measurement capabilities of PMUs and the baseline techniques used to extract information using a cache side channel, we could assert that PMUs are the perfect candidates to build a detection mechanism for this kind of attack. While several works in the literature have followed this path, a recent result has argued that micro-architectural level information obtained from HPCs cannot distinguish between benignware and malware [63]. Similarly, another work has illustrated why many of the results in the literature cannot be considered reliable [14].

In this article, we come back to this problem and try to capture some common features which can be used to define *detection metrics* based on measurements obtained through HPCs. We use these metrics to detect whether a process running in the system is carrying out an attack—independently of whether the attack is carried out after having exploited some transient execution CPU vulnerability.

Our detection mechanism is system-wide. In this sense, we do not make any assumption on which process is the attacker and which is the victim. We also directly account for scenarios where multiple linked processes are used to mount the attack (e.g., relying on the `fork()` system call). We exploit the information gathered at runtime to deem some processes as *suspected*. In more detail, we concentrate on detecting the usage of side channels to extract information during an attack to indicate the *possibility* that the process is malicious.

We explicitly acknowledge that our detection mechanism is fallible due to the degree of uncertainty associated with this kind of mechanism. Therefore, we do not take any destructive action with respect to the running process. Instead, we couple our detection capabilities with mitigation actions. We propose different mitigation actions automatically enforced by the operating system as soon as a process is suspected as malicious. They entail a limitation in the scheduler freedom at deciding what CPU resources should be assigned to some process or the selective (per-process) activation at runtime of security patches against transient execution vulnerabilities.

To reduce the incidence of false positives and negatives, we rely on a self-adjustable observation window coupled with a scoring system. This approach is meant to reduce the probability that benignware with pressure on the memory hierarchy is suspected or to increase the likelihood to suspect processes that perform many non-malicious actions before carrying out the attack.

Our detection mechanism and the mitigations mentioned above have been implemented at kernel-level in Linux and have been exercised on multiple processors of the x86 family. We have used our patched kernel for a month, also in daily usage [1] —the patched kernel has also been used while typesetting this paper. No false-negative has been observed under that daily usage workload. Of course, this is not a guarantee that our approach could be used to enforce more intrusive policies for suspected processes like, e.g., killing suspected processes. Instead, it is an indication of the viability of using HPCs as building blocks for articulated detection mechanisms and for devising strategies where the setup of security-oriented patches can be put in place on a dynamic and per-process basis—rather than paying the cost of these patches by default when any process is active. Our reference implementation is released as open-source software[2].

We finally compare the performance penalty introduced in the system by these different mitigation strategies relying on standard benchmarks for operating systems [34].

Overall, our main contributions can be summarized as:
- we introduce a practical mechanism to build metrics from measurements obtained from HPCs to detect that some cache side channel is currently being used;
- we propose an observation window and a scoring system to reduce false positives and negatives when considering a process as suspected;
- we propose a system-wide detection approach for userspace applications, which makes no assumption on which process is the victim and which is the attacker;
- we enable per-process or per-CPU mitigation strategies;
- we describe a reference implementation in the Linux kernel;
- we show the effects on performance of our proposal on multiple generations of Intel CPUs, using standard benchmarks.

The remainder of this paper is structured as follows. In Section 2 we discuss the threat model that we have considered. Section 3 discusses related work. The metrics used in our detection system are introduced in Section 4. The system-wide implementation of our detection mechanism is presented in Section 5. We discuss our mitigation strategies in Section 6. The experimental assessment of our proposal is provided in Section 7. Section 8 concludes the article.

---

[1]A video demonstration of the operations of our detection mechanism is available at https://youtu.be/XGQ4TuqtTAI.
[2]The code is available at https://github.com/HPDCS/linux-detection.

## 2 THREAT MODEL

We consider an attacker trying to carry out a cache-based attack and extract information from a co-located victim on the same platform. The attacker is thus sharing some architectural components with the victim, such as the First-Level Cache (L1) [30, 49] or the Lowest-Level Cache (LLC) [20, 27, 28, 37]. In the most general setting, the victim can be some userspace process, a virtual machine in a multi-tenant cloud environment, or the underlying OS kernel. We do not make assumptions on the privileges with which the attacker is running, nor on whether the side channel is being used to extract leaked information *after* some transient execution attack has been carried out. Indeed, as mentioned, we are interested in detecting the usage of a cache side channel to extract information while the attack is in progress. This allows us to detect also popular attacks such as Meltdown [36], Spectre [33], and Foreshadow [52, 58].

Concerning transient attacks, we assume that security mitigation patches such as KPTI are not necessarily active (hence the attack is not prevented) but are available in the compiled operating system binary. Indeed, we propose a mitigation mechanism that allows to selectively re-enable these patches on a per-process basis, just to prevent the attack, if a process is suspected as malicious, as we shall describe in Section 6.

We also assume that the operating system's kernel is not compromised in any way. In particular, we assume that any data acquired by the kernel is not tampered with by an attacker and that the routines executed by the kernel are similarly not altered by any attacker. Hence, in our proposal, we assume that no attack is run from kernel space, e.g. the victim has not loaded any malicious kernel module that would mount a side-channel attack. The operating system's kernel internal and external security is an orthogonal security aspect to the proposal discussed in this article.

## 3 RELATED WORK

We can relate our proposal to two different families of countermeasures to attacks, namely *detection* and *prevention*. On the detection side, using HPCs is not a new idea. Many works in the literature have relied on HPCs for this purpose, e.g., for exploit detection [50, 60, 61, 64], malware detection [15, 17, 29, 45, 48, 55], firmware verification [56, 57], integrity checking [9, 39], or vulnerability analysis [12]. Unlike our proposal, these works mainly cope with attacks not explicitly oriented to cache side channels, like ROP (or more generally control flow tampering) or similarity-based malware detection.

We share the goals with a set of works that rely on HPCs to detect side-channel attacks [6, 11, 26, 40, 43, 46]. In general, these approaches rely on machine learning mechanisms, concentrate on specific attacks, do not support system-wide detection, require to know beforehand what the attacking process is, or do not consider the possibility of relying on selectively activated software patches. These are all major differences from our proposal.

On the prevention side, an important mitigation for the Meltdown attack is KPTI [18]. When running in user mode, this mitigation strategy drops the historical sharing of the kernel-level virtual-to-physical translation metadata (namely, the kernel-level page tables). In user mode, a process only observes a minimal amount of data and code belonging to the kernel, i.e., the data and code, to allow a safe transition to kernel mode upon interrupts and system calls. This minimal set of code, when activated, performs a page-table switch, which allows accessing the whole virtual address space of the operating system kernel. Before returning to user space, the user-land trimmed page table is put back in place. All major operating systems have adopted this scheme.

The main problem with this approach is that, for it to work correctly, a page table change must be accompanied by a flush of all virtual-to-physical translation entries in the caches—some of these are done automatically when updating the page-table pointer, e.g. CR3 on x86 CPUs. This incurs additional runtime costs, which have been quantified to be up to 30% of the execution time observed when this mitigation is not in place [21].

Concerning Spectre attacks, CPU vendors have introduced hardware mitigations for Speculative Store Bypass (SSB) [3, 24]. Examples are the Indirect Branch Restricted Speculation (IBRS) mitigation, which restricts speculation of indirect branches, the Single Thread Indirect Branch Predictors (STIBP) mitigation, which prevents indirect branch predictions from being controlled by the sibling hyperthread, and the Indirect Branch Predictor Barrier (IBPB) mitigation, which ensures that earlier code's behavior does not influence later indirect branch predictions. The software community has been cold towards these mitigations, as they have been reported to slow down typical workloads up to 50% [13].

On the other hand, one software mitigation to Spectre-like attacks, which introduces a minimal overhead, is the retpoline [1]. It is a software construct that ensures that if the CPU is mispeculating due to some attack being carried out against some branch prediction unit, then the pipeline will be filled with an infinite loop. This prevents arbitrary code execution, which could also induce data leaks.

Our work grounds on these and other mitigation patches and strategies. However, our goal is to enable these patches at a very fine grain, i.e., whenever a process is suspected as malicious. This finer grain should reduce the performance impact in the general lifetime of the system while ensuring a higher security level with respect to an utterly unpatched system.

## 4 DETECTING SIDE-CHANNEL ATTACKS

The overall architecture and methodology that we use to enable prompt detection of side-channel attacks are depicted in Figure 1. We rely on a combination of measures taken from HPCs in real-time, which allows us to discriminate processes that are more likely to perform operations on the cache hierarchy, indicating that they are mounting a side-channel attack. At the kernel level, we have four major components involved in the system-wide monitoring of the attacks to detect the activity of malicious processes. The *Monitor* module directly interacts with hardware performance counters, programming them to acquire the measures to build our detection metrics. Data coming from HPCs are stored directly in a process' `task_struct`. The *Detector* module relies on these data to compute detection metrics and deem a running process as suspected or not—again, this information is stored in the task struct. If a process is suspected, the *Mitigator* module will detect it and apply proper mitigations. The *Scheduler* module interacts with the operating system's scheduler. It is one of the fundamental components to enable system-wide detection and per-process mitigations: every time that a different `task_struct` is scheduled, both the *Mitigator* and the *Monitor* modules are notified to enable/disable mitigations and reprogram HPCs, respectively, to account for the newly scheduled process.

### 4.1 Architectural Details and Preliminary Work Hypotheses

Since we are interested in the malicious usage of caches to extract information, it is beneficial to discuss how the CPU interacts with the caching subsystem briefly. Its organization is depicted in Figure 1 for the Intel architecture, which we use as our reference, where there are three levels of CPU caches. The caches closer to the CPU are smaller and faster, and the caches further away are larger and slower. At the first level, there are two caches, L1i and L1d, which keep code and data, respectively. The L2 cache unifies code and data and, in almost every x86 CPU, represents the last cache level private to the core, while, as the final level, there is the Last Level Cache (LLC), a shared memory level among all cores of the same die. Two important cache properties to be considered are *inclusiveness* and *associativity*. The former defines the way a cache level behaves with respect to the higher ones, which can be:

(1) inclusive: this level always contains data stored in the higher levels;
(2) exclusive: precisely the opposite of the previous one;
(3) non-inclusive: it does not guarantee that higher levels state is a subset of the current one.

This work considers inclusive caching systems since they represent the most diffused chipset for Intel processors. On the other hand, associativity is a strategy that divides a single cache level into multiple sets, where part of the physical address is used to index into the corresponding cache set. It is helpful to reduce chip complexity while providing a more efficient cache implementation. Also, the page table walk firmware relies on the CPU caches to further improve the performance of a TLB miss [53]. In particular, the PTEs for the different page table levels are not stored only in the CPU caches, but modern processors also store them in page table caches or translation caches [5]. Independently of the associativity strategy, we work at the granularity of the single cache line.

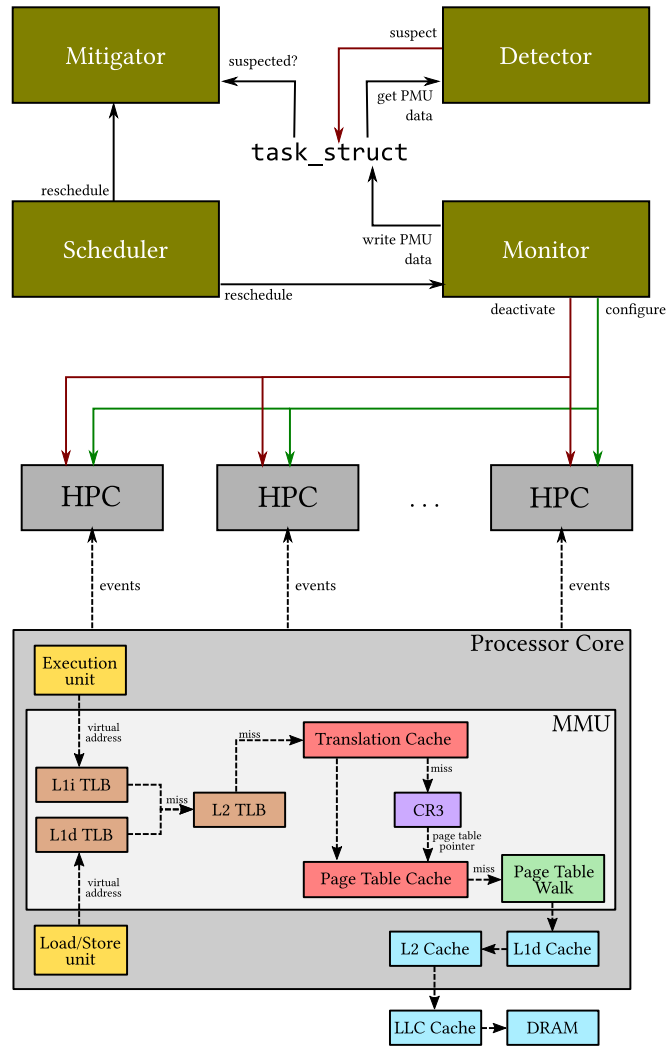

Fig. 1. Overall detection architecture and components involved in memory access—hit paths are not shown.

HPC operations are based on the concept of *hardware events*. Examples of hardware events are a write operation in memory, a cache miss, or the fact that a conditional branch in the execution flow of the program has been taken. The events that a processor is aware of (and can be measured) are highly coupled with the actual processor architectural family. This is because the generation of a hardware event is physically triggered by data paths or control signals implemented in the actual control unit of the CPU, which is often subject to partial or complete re-implementation across different families of processing units. In the literature on hardware profiling, this extremely low-level information is often referred to as *micro-architectural events*.

Each HPC can be configured to track one micro-architectural event at a time. The software interface to control an HPC is implemented by relying on a couple of *model-specific registers* (MSRs): (1) a *selector* (or *control*) register which is used to specify the HPC operating mode and the micro-architectural event to be observed; (2) a *counter* register, which is incremented every time the associated micro-architectural event is triggered—of course, the counter can overflow.

The *control* register can be used to activate two different operating modes. When the HPC is configured to work in *counting mode*, the *counter* register simply accumulates the number of target micro-architectural events observed while any program is running on the CPU. On the other hand, when the HPC is working in *sampling mode*, the system generates a Performance Monitoring Interrupt (PMI) as soon as the *counter* register overflows. We have configured our reference implementation to rely on sampling mode, as we shall discuss.

## 4.2 Detection Metrics

In this Section, we describe the metrics we have devised to determine whether some malicious side-channel activity is going on. In an initial set of metrics, we relied on the idea that the exploitation of a side channel is based on bringing the caching system into a known initial state. Successively, the attacker attempts to determine whether some change has occurred in the cache state. However, considering inclusive caching systems, which represent our target, we know that bringing the cache into a given state (e.g., a cache line is flushed or a cache set is primed) means performing an operation that is necessarily reflected into the state of caches at all the levels, from L1 to LLC.

Based on this observation, we decided to relate to each other volumes of micro-architectural events that are generated at different levels within the caching system. At the same time, we wanted to focus on events that are not easily manipulable (in terms of their volume generation at a specific cache level) by an attacker[3]. Therefore, we decided to avoid considering cache hits and to focus exclusively on cache miss events.

Independently of whether the initial state of the cache when the attack is started is based on cache-line flushes or cache-set primes, a side channel is anyhow based on re-accessing the same cache line to discover changes into the state. Clearly, the "interface" for the observation is the L1 cache, but the actual cache access needs to pass through lower levels if a miss is observed. On the other hand, if a miss has been experienced at the L1, the likelihood of observing misses at the lower levels is expected to be high. In fact, by the nature of a cache side channel, the victim either brought some cache line into the caching system, up to L1, or gave rise to cache line replacement involving all the cache levels because of inclusiveness.

Considering that in inclusive caching systems the volume of cache misses at upper levels is greater than or equal to the ones at the lower levels, our first two detection metrics are based on the ratio between the number of cache misses at L1 denoted as $L1_{miss}$, and the corresponding values at lower levels denoted as $L2_{miss}$ and $LLC_{miss}$. We have therefore two predicates $\mathcal{P}_1$ and $\mathcal{P}_2$ for building our side channel suspicion, which are based on relating the aforementioned ratios to thresholds, namely:

$$\mathcal{P}_1 : L2_{miss}/L1_{miss} > \phi_1 \qquad (1)$$

$$\mathcal{P}_2 : LLC_{miss}/L1_{miss} > \phi_2 \qquad (2)$$

with the values of $\phi_1$ and $\phi_2$ both included in the interval [0,1]. Clearly, the value zero for these thresholds leads to a highly conservative setting where the predicates always hold, leading to suspicion independently of the actual execution pattern. Values closer to one are more representative in terms of the ability to discriminate between malware and benignware.

---

[3]As an example, an attacker might easily give rise to volumes of cache hits at the L1 in an uncorrelated manner to the cache hits observable at the LLC.

Another interesting point about caching is that the caching hierarchy typically supports data prefetch in order to implement anticipated reads useful to serve access locality by the applications. For example, this is the case of the L2 cache in Intel processors. However, a cache side channel is typically based on activities that target a specific cache line. Hence, we may expect that prefetched data may result useless. On the other hand, the scarce exploitation of prefetched data is challenging to be discovered by relying on miss events. For this reason, we devised an additional metric based on the relation between the number of write-back operations for cached lines at the L2 cache, which we denote a $L2_{write-back}$ and the number of lines fetched (including the prefetched ones) still at the L2, which we denote as $L2_{lines-in}$. Accordingly, we derive a third predicate $\mathcal{P}_3$, based on an additional threshold $\phi_3$, still having a value in the interval [0,1], in order to determine the side channel suspicion, according to the following expression:

$$\mathcal{P}_3 : L2_{write-back}/L2_{lines-in} < \phi_3 \tag{3}$$

Essentially, predicate $\mathcal{P}_3$ is intended to capture all the scenarios where data update activities do not comply with locality expectations (especially for very low values of $\phi_3$), which can be an indication of some unexpected non-local behavior.

The above-described metrics and predicates are tailored at direct cache side-channel attacks, namely those attacks that are based on managing cache lines/sets explicitly with data in the address space of the attacker. Another way of attacking the cache to mount a side channel is to have indirect attacks based on the fact that memory management metadata, in particular, page table entries, are still cached (see Figure 1). This may lead to evict cache sets with these metadata, thus enabling the determination of the metadata re-access time to discover whether some victims had conflicting accesses to the same cache line used to keep the page table entries. To cope with these kinds of attacks, we devised an additional metric, based on the number of TLB misses at the second level of the page walk, denoted as $TLB_{miss-level-2}$ and the number of L1 cache misses. In particular, a high value of the ratio between TLB misses and L1 misses is representative of a behavior not conforming with classical locality (namely, a behavior not conforming with good exploitation of already carried out virtual-to-physical address translations). This may therefore be a behavior where a cache miss is generated just because of the will to fill the TLB (upon a TLB miss) with data leading to a cache line replacement. In order to determine the side channel suspicion in such indirect attack scenarios, we have therefore the following additional predicate:

$$\mathcal{P}_4 : TLB_{miss-level-2}/L1_{miss} > \phi_4 \tag{4}$$

where $\phi_4$ is this time not constrained in the interval [0,1].

At this point, we can combine the above-defined predicates to determine whether to suspect that a side channel exploitation is taking place, or not. Before doing this, for direct side-channel attacks, we exploit again the ratio $TLB_{miss-level-2}/L1_{miss}$ to define the following additional predicate:

$$\mathcal{P}_5 : TLB_{miss-level-2}/L1_{miss} < \phi_5 \tag{5}$$

with $\phi_5 < \phi_4$. Actually, $\mathcal{P}_5$ expresses the fact that there is no bias generated by a direct side-channel attack in terms of the increase in the volume of TLB misses, with respect to the volume of L1 misses related to actually-accessed data. In fact, direct attacks only exploit data in the address space (not memory address translation metadata). Overall, for direct attacks we define a combined predicate $\mathcal{S}_1$ to determine whether to raise a suspect as the following combination of $\mathcal{P}_1$, $\mathcal{P}_2$, $\mathcal{P}_3$, and $\mathcal{P}_5$:

$$\mathcal{S}_1 = \mathcal{P}_1 \wedge \mathcal{P}_2 \wedge \mathcal{P}_3 \wedge \mathcal{P}_5 \tag{6}$$

For indirect attacks, we just have $\mathcal{P}_4$. Finally, a side channel suspicion is raised based on the following combination of $\mathcal{S}_1$ and $\mathcal{P}_4$:

$$\mathcal{S} : \mathcal{S}_1 \vee \mathcal{P}_4 \qquad (7)$$

### 4.3 Setting up the thresholds

Basing the detection on measures from HPCs compared against thresholds has already been identified in the literature as a possible pitfall [14]. However, a significant identified issue is related to how these thresholds are set and how they are employed. In particular, while it is clear that thresholds could be circumvented (e.g., by inducing page faults to affect the accuracy of the measured events [14]), we emphasize that we play on the safe side. Indeed, we use thresholds to discriminate between *relations* among events, which are in any case representative of the ultimate utilization of the side channel to extract information. Furthermore, our approach to side channel detection is based on combining metrics (via the combination of predicates involving these metrics), which should favor robustness.

Nevertheless, relying on hardcoded thresholds would make the approach difficult to maintain over time, requiring significant manual intervention. Changes in the hardware, or peculiarities of specific CPUs, are some of the aspects which could require to re-tune the thresholds.

The behaviors which we discriminate with our metrics depend mainly on the architecture of the cache of the machine in which the thresholds are used to discriminate a process as malicious or not. To this end, our system explicitly allows defining the values for the different thresholds $\phi_i$ for the actual machine on which we perform the detection at configuration time. In our reference implementation—see Section 7 for additional details—this is done by running:

- A set of side-channel attacks in a controlled environment.
- A set of benchmark applications from different fields.

By relying on these attacks and on the behavior of the benchmarks (which represent the benignware part), we can estimate proper values for the thresholds, which are used in our detection mechanism. In particular, we define a threshold value as the average of the two (already averaged values) for the cases of the run attacks and benchmarks. We note that including benignware execution in the setup of the thresholds gives rise to a somehow conservative estimation of the threshold values that is, anyhow, not unfavorable to non-malicious software.

We note that to avoid bias in the experiments, the synthetic attacks which we carry out at system startup are different from the ones which have been used to test our approach. This is an approach similar in spirit to techniques that perform preliminary probing of the hardware architecture in order to carry out an attack effectively [54].

## 5 SYSTEM-WIDE DETECTION AND REFERENCE IMPLEMENTATION

As mentioned, our goal is to carry out a system-wide detection of possible attacks relying on side channels to extract information. This detection is carried out at the kernel level—our reference implementation is based on a set of patches applied to Linux 5.4.145. In our implementation, we have targeted the Intel architecture, considering its widespread nature [47] and the fact that it has been repeatedly subject to multiple attacks in the last years. Nevertheless, as we discuss, our reference implementation can be easily ported to other architectures, such as AMD.

### 5.1 Selected Monitoring Events and Strategy to Acquire HPC Data

We must first give additional details on how we have configured HPCs. Sampling has been set to follow the number of clock cycles—CPU_CLK_UNHALTED on Intel CPUs, PMCx076 (*CPU Clocks not Halted*) on AMD. With respect to the measures, we have tried to select stable measurements to instantiate the proposed metrics. In particular, the following events have been selected [4, 23]:

- $L1_{miss}$ is mapped to the `L2_RQSTS.ALL_DEMAND_DATA_RD` event. On AMD, a suitably corresponding event is PMCx041 (*Data Cache Misses*);
- $L2_{miss}$ is mapped to the `L2_RQSTS.DEMAND_DATA_RD_MISS` event. On AMD, a suitably corresponding event is PMCx07E (*L2 Cache Misses*);
- $LLC_{miss}$ is mapped to the `OFFCORE_REQUESTS.L3_MISS_DEMAND_DATA_RD` event. On AMD, a suitably corresponding event is PMCx0E0 (*DRAM Accesses*);
- $L2_{write-back}$ is mapped to the `L2_TRANS.L2_WB` event. On AMD, a suitably corresponding event is PMCx07F (*PMCx07F L2 Fill/Writeback (L2Writebacks bit set)*);
- $L2_{lines-in}$ is mapped to the `L2_LINES_IN.ALL` event. On AMD, a suitably corresponding event is PMCx07F (*PMCx07F L2 Fill/Writeback (L2Fills bit set)*);
- $TLB_{miss\_level2}$ is mapped to the `DTLB_LOAD_MISSES.MISS_CAUSES_A_WALK` event. On AMD, a suitably corresponding event is PMCx045 (*PMCx046 Unified TLB Miss*).

We have configured our implementation for scenarios where Simultaneous Multi-Threading (SMT) is disabled. This choice is motivated by the fact that, with SMT disabled, Intel CPUs of various generations offer at least eight programmable HPCs, which are enough to sample all the parameters involved in our metrics. With SMT enabled, this number would be reduced to four on many CPUs. Furthermore, by disabling SMT, we remove the noise in the experimental assessment related to the need to time-share the HPC units to gather data related to different measures, thus focusing better on the validity of our approach. Using our approach with SMT enabled is possible, but it requires techniques to share HPCs to read multiple measures, which are out of the scope of this article.

In our reference implementation, we have tackled the cost of running the PMI handler by installing a custom interrupt handler lined up on a free vector in the Linux Interrupt Descriptor Table (IDT). This custom interrupt handler, which is reserved for PMIs, bypasses the traditional activation scheme for interrupt management in Linux. Indeed, Linux typically manages a hard interrupt by activating multiple nested functions, in particular related to the identification of the proper Interrupt Service Routine in charge of managing the IRQ. This is a cost that cannot be paid to just record a number of occurred events from an HPC.

Our custom stub accounts for the bare minimum amount of actions required to serve the interrupt request (namely: possibly execute `swapgs`, change the page table if KPTI is active, set the per-CPU flags used to determine the execution in kernel mode, take a CPU snapshot). After the actual mode change, we filter out possible spurious interrupts, and we collect samples from HPCs. We then compute our detection metrics[4], determine whether to consider the current process as suspected or not, and finally return from interrupt. The obtained data are saved on a per-process basis in the `task_struct` of the thread currently running on the core, which is serving the interrupt request.

Another important aspect is related to the fact that HPCs are shared among processes scheduled on it. As shown in [14], underestimating this property leads to an inconsistent behavior of the system-wide detection mechanism. To cope with this aspect, upon context switch (`prepare_task_switch()`), if the observed data is enough, we early evaluate the metrics for the about-to-be-descheduled process. On the other hand, before returning control to the newly-scheduled process (namely, in `finish_task_switch()`), we logically reset the HPCs and start the measurement of the about-to-be-scheduled process. In this way, we do not mix HPC data coming from the execution of different processes if a thread is scheduled/descheduled in the middle of an observation window, which could lead to an erroneous detection.

---

[4]We have implemented metrics evaluation in integer arithmetic, both to reduce the execution time and not to poison the FPU—we are not explicitly saving the FPU state.

As a last note, we have configured HPCs to explicitly filter out activities when running in kernel mode—the USR configuration bit in the HPC control register. In this way, every time that we run in kernel mode (also to extract the values of some HPCs upon a PMI), we do not overcount the measures taken from HPCs—this solves another source of unreliability observed in [14][5].

## 5.2 Observation Windows

As discussed, an effective system-wide detection requires filtering out all the activities not directly related to the instruction sequence of the attack to avoid pollution in the observed data. Given that HPCs cannot leave the micro-architectural domain, it is impossible to identify program phases just by counting low-level events. This aspect could allow an attacker to blend the malicious code into any program, concentrating the attack phase to a limited execution time window.

To cope with this problem, we divide the entire observation period into *time slots*, which are handled as observation windows of HPC values that are inspected one by one. This allows discriminating among different execution phases. Such a discretization is applied to the number of *elapsed clock cycles* (which defines a constant unit among all running processes) rather than events such as *retired instructions*—they may warp the time slot depending on the executed instruction[6]. This window is preserved across context switches and is not shared among processes/threads, thus guaranteeing a coherent inspection of the execution flow. In other words, if a process/thread is descheduled in the middle of an observation window, once it is rescheduled, we resume collecting HPC data from the same exact "point" in the observation window at which it was descheduled. We also note that this approach allows overcoming the problem affecting other works (see, e.g., [41]), in which data collection is associated with the entire program execution.

The related HPC's overflow defines the beginning and end of a time window. It is essential to determine the time slot size so that the observable data is enough to discriminate meaningful program phases. A too-small size may cause each slot to provide noisy and poor information, while a too-large one will eventually fall into the same pitfall as in [41], i.e., too much-aggregated data. Furthermore, the size of the time window is directly related to the overhead that the detection architecture introduces in the system because smaller slots imply more interrupts to be processed.

Similarly to what we have discussed in Section 4.3, we determine the minimum and maximum thresholds for the observation window at system startup, guaranteeing stable measurements. This is done via an adaptive approach: if we observe a large fluctuation in the data observed across two consecutive windows, we reduce the size of the window (up to a compile-time defined minimum threshold, which accounts for the overhead in the measurement). Conversely, if variations are minimal, we increase its size (up to another compile-time defined maximum).

## 5.3 Suspecting Malicious Processes

After calculating the metrics in the PMI, they are compared to the respective thresholds, thus determining if the predicates driving suspicion hold—see expression (7). Based on the inequalities results, we deem a process as malicious or not. Obviously, the classification of a process cannot be made based on a single observation because we would have an excessive number of false positives considering that, during its execution, a process can assume different behaviors. For this reason, we have introduced a scoring system. The process's score will vary during execution as follows:

- the score is increased by $\alpha$ if the results of the comparison between metrics and thresholds show a behavior similar to a side-channel attack;
- the score is decremented by $\beta$ if the metrics do not detect any abnormal situation.

---

[5]Unfortunately, Intel confirmed [25] that using CPL through the PMC's OS-USR bits may lead to an incorrect result, such that the sum of OS-data and USR-data is not equal to the result obtained by counting without filters. This phenomenon is probably due to the high out-of-order execution degree, which makes it hard to associate the $\mu$op execution with the correct execution ring near a mode transition. Nevertheless, for our specific context, we have performed some tests in order to quantify this error and observed very minimal error values (less than 0.1%).
[6]Every instruction requires a certain number of clock cycles to be carried out, which varies according to several factors (e.g., the memory state).

If the score reaches the value of a threshold $\gamma$, then the process becomes suspected. $\alpha$, $\beta$, and $\gamma$ are tunable hyperparameters of our model. These parameters are related to each other in the following way. $\alpha$ indicates how fast a process becomes suspected: the higher the value, the smaller is the number of positive evaluations of the metrics required to flag it as malicious. Conversely, $\beta$ determines how fast a process that was (incorrectly) considered suspicious starts again to be deemed benign. $\alpha$ and $\beta$ can be therefore used to control the responsiveness of our scoring system towards punctual activities (i.e., possibly malicious or not) exhibited within an observation window. In the general case, we assume $\alpha \geq \beta$ to allow for a prompt-enough detection of a malicious process. Conversely, $\gamma$ directly controls when a process becomes flagged as malicious. To some extent, it indicates the amount of data that the system tolerates to leak before deeming a process as suspected. In Section 7, we provide an empirical assessment of the behavior of our approach with respect to these parameters.

Once a process is suspected, this information is stored into `current->mm->flags`—we exploit bit 27, which is currently unused. We have explicitly decided to rely on the `flags` field in the `mm` data structure because, upon a `fork()`, this data structure is automatically copied by the kernel, to make it inherited by the child process. In this way, also if the attacker tries to jeopardize our detection system by relying on a multi-process attack, the behavioral information associated with children and the parent processes is shared.

## 6 MITIGATION STRATEGIES

Our kernel-based detection subsystem can flag a process as suspected. A suspected process is one for which we can implement mitigations. We note that this is not a destructive operation: even if we have incurred a classification error (i.e., a false positive), the fact that we enable mitigations will not cause runtime errors (e.g., abnormal termination) in the wrongly-suspected process. Indeed, we could only cause a performance slowdown. Nevertheless, considering the overall system, this slowdown will not be comparable to that observed if the mitigations we discuss here were activated by default for all processes—see Section 7 for the overhead assessment. We have foreseen two families of mitigations: one related to side-channel attacks in general and one pertaining to transient execution vulnerabilities. The mitigations we put in place have value independently of whether our approach is used to detect the attacks or other support would be used to determine (potentially) malicious processes.

### 6.1 Side-channel Attack Mitigations

Multiple mitigations belong to this family. The first one entails that, in `finish_task_switch()`, before returning control to a thread of a suspected process, we flush the last-level CPU cache. This ensures that no data from other processes is available in the cache to be leaked. Of course, this is an intrusive operation performance-wise. However, it mainly affects the execution of the suspected process, for which the cache must be again warmed up upon its reschedule[7].

A second mitigation we devised tries to mitigate the fact that an attacker is likely running on a CPU core that is "close" to the cache used by the victim. Therefore, an additional strategy is to change the affinity of the attacker to move it to a different core which is not sharing the same level of cache with the victim. We note that this mitigation could also be performance-intrusive, particularly for applications that have explicitly set their affinity, e.g., to control their memory-access latency on NUMA machines. However, the system administrator always can change the affinity for any thread. Hence our approach mimics such a kind of housekeeping job, in this case carried out for security purposes.

---

[7]We think that the performance penalty paid by the OS kernel when managing interrupts that do not find cached data after the reschedule of the suspected process can have a limited impact, with respect to the fact that upon the reschedule the CPU-core is anyhow devoted to the specific activities related to the execution flow of the suspected process.

## 6.2 Transient Execution Mitigations

Another mitigation that we explicitly put in place is per-process enabling of KPTI. The baseline implementation of KPTI in Linux has been slightly changed to support this mitigation.

In particular, while we maintain the order-1 allocation (8 KB) for the first-level page table (pointed to by the CR3 register), which allows having two different views of the address spaces for each process (one for user mode, one for kernel mode), by default all processes rely on only one order-0 page table, which maps the whole kernel address space. This configuration resembles the traditional organization of the memory map in Linux before the introduction of KPTI. The two first-level page tables are kept synchronized following a scheme that resembles the one currently adopted to synchronize them whenever a userspace application allocates new physical memory. In particular, every time that a new set of physical pages is allocated, the kernel with KPTI enabled invokes `__pti_set_user_pgtbl()` which materializes in the user-level page table the newly-allocated virtual-to-physical translation metadata (the page table chain), also explicitly setting the NX bit, if available on the current architecture. This same scheme is adopted upon a `fork()`. We retain this scheme, although we explicitly differentiate between the user- and the kernel-level page table—this distinction is somewhat implicit in the current standard implementation of KPTI and relies on some hardcoded macros.

Upon a mode switch, the `switch_to_kernel_CR3` macro is used by the kernel to open access to the whole address space of the kernel. Upon return to userspace, the `switch_to_user_CR3` macro returns to the user-level page table. This scheme is done every time the machine transitions from user to kernel mode and vice-versa. Our goal is to selectively activate this scheme in a per-process way, reducing at most the cost for this operation.

To this end, we recall that a process becomes suspected while running in kernel mode, namely while a PMI is being processed. In that case, we set a flag in `current->mm->flags`. When returning to user mode, we explicitly check this flag. If it is set, we invoke `switch_to_user_CR3`. This is enough to start applying the patches for a suspected process. Conversely, we cannot check this flag when transitioning from user to kernel mode. This is because we do not have access to `current`, which is stored in per-CPU variables, which are not accessible if the user-mode page table is set.

To check if we have to invoke `switch_to_kernel_CR3`, we exploit the fact that the two first-level page tables belong to an order-1 allocation and are therefore contiguous (both in the virtual and in the physical address space). We have inverted the user and the kernel page table with respect to the current implementation of KPTI. This means that the user page table follows the kernel page table. Given the contiguousness of the pages, it is sufficient to check if bit 13 of the address contained in CR3 is set to 1. If this is the case, the thread enters kernel mode with the user-mode page table. This means that the thread belongs to a suspected process, and we, therefore, have to invoke `switch_to_kernel_CR3`. On the other hand, if the bit is cleared, the process is not suspected, and the whole kernel virtual address space is already visible.

Of course, we want to account for suspected multi-threaded applications explicitly. In this scenario, two threads could be concurrently running on multiple CPUs. We want to minimize the time window when a thread is running with patches enabled, and another is not. As mentioned, to activate the patch, a thread belonging to a suspected process must perform a mode switch from kernel mode. To this end, after flagging a process as suspected, we explicitly send an Inter Processor Interrupt (IPI) to all other cores. This operation will require all CPU cores to transition to kernel mode. In this way, if a thread of the suspected process was running, the mode change will result in patch enabling.

Table 1. An overview of considered cache side-channel attacks and references to the used implementations.

| Name | Same-Core | Cross-Core | Shared Memory | Measurement |
|---|---|---|---|---|
| Evict + Time [35] (taken from [8]) | ✓ | ✓ | ✗ | time |
| Prime + Probe [27, 37] (taken from [9]) | ✓ | ✓ | ✗ | time |
| Prime + Abort [16] (taken from [9]) | ✗ | ✓ | ✗ | TSX |
| Flush + Reload [59] (taken from [9]) | ✓ | ✓ | ✓ | time |
| Flush + Flush [19] (taken from [9]) | ✓ | ✓ | ✓ | time |
| Xlate + Time [53] (taken from [9]) | ✓ | ✓ | ✓ | time |
| Xlate + Probe [53] (taken from [9]) | ✓ | ✓ | ✓ | time |
| Xlate + Abort [53] (taken from [9]) | ✓ | ✓ | ✓ | TSX |
| *Meltdown* (taken from [10]) | | | | |
| *Spectre* (taken from [11]) | | | | |
| *Foreshadow* [52, 58] (taken from [8]) | | | | |

A similar mechanism has been put in place to enable/disable several other mitigation techniques, namely: i) Microarchitectural Data Sampling (MDS); ii) Spectre v1, v2, L1TF mitigations; iii) SSB mitigations; iv) KVM Non-Executable Huge Pages; v) TSX Asynchronous Abort. This is supported by quickly checking the flag in `current->mm->flags` to determine whether one specific mitigation should be activated, which might, in turn, require modifying the content of some MSR value (as in the case of SSB mitigations).

## 7 EXPERIMENTAL ASSESSMENT

### 7.1 Experimental Setup

We have carried out an experimental assessment relying on multiple generations of Intel CPUs, namely using the following processors:

- i7-6700HQ 4x (SMT) L1 64KB (I,D) 8-way, L2 256KB, shared L3 6MB 12-way;
- i7-7600U 2x (SMT) L1 64KB (I,D) 8-way, L2 256KB, shared L3 4MB 16-way (with TSX);
- i5-8250U 4x (SMT) L1 64KB (I,D) 8-way, L2 256KB, shared L3 6MB 12-way;
- i7-9750H 6x (SMT) L1 64KB (I,D) 8-way, L2 256KB, shared L3 16MB 16-way;
- i7-10750H 6x (SMT) L1 64KB (I,D) 8-way, L2 256KB, shared L3 12MB 16-way.

To set up the thresholds and observation windows used by our detection system, we have run versions of the attacks listed in Table 1, as well as the following set of benignware applications: (1) Firefox, with both textual page, multimedia content access, and browser benchmarks such as JetStream2; (2) VLC, with both large and short videos and random skip of video portions, as well as repositioning; (3) Evince Reader, with both small and large size pdf files, and random skip of pages; (4) gedit for editing textual files of different sizes and random positioning onto the file portion to be edited; (5) all the kernel-level threads operating within the Linux kernel.

---

[8]https://github.com/vusec/revanc.
[9]https://github.com/vusec/xlate.
[10]https://github.com/paboldin/meltdown-exploit.
[11]https://github.com/Eugnis/spectre-attack.

Table 2. Comparison between HPCs and Software Instrumentation on all the architectures. Err represents the distance (%) between HPCs and SW while HPCvar shows the HPCs variation coefficient.

|          | **i7-6700HQ** | | | | **i7-7600U** | | | | **i5-8250U** | | | |
|----------|------|------|------|--------|------|------|------|--------|------|------|------|--------|
|          | HPCs | SW   | Err  | HPCvar | HPCs | SW   | Err  | HPCvar | HPCs | SW   | Err  | HPCvar |
| loads    | 4494K | 4744K | 5.2% | ~0%   | 4494K | 4744K | 5.2% | ~0%   | 4494K | 4744K | 5.2% | ~0%   |
| L1 miss  | 400K | 308K | 29%  | 2.0%   | 515K | 308K | 66%  | 2.9%   | 521K | 308K | 69%  | 1.6%   |
| L3 miss  | 8557 | -    | -    | 6.4%   | 7798 | -    | -    | 3.9%   | 6539 | -    | -    | 3.6%   |
| L2 lines | 185K | -    | -    | 7.9%   | 221K | -    | -    | 3.1%   | 224K | -    | -    | 4.7%   |
| TLB miss | 9168 | -    | -    | ~0%   | 9382 | -    | -    | 5.7%   | 8981 | -    | -    | 1.9%   |

|          | **i7-9750H** | | | | **i7-10750H** | | | |
|----------|------|------|------|--------|------|------|------|--------|
|          | HPCs | SW   | Err  | HPCvar | HPCs | SW   | Err  | HPCvar |
| loads    | 4494K | 4744K | 5.2% | ~0%   | 4495K | 4744K | 5.5% | ~0%   |
| L1 miss  | 513K | 312K | 64%  | 2.2%   | 517K | 308K | 67%  | 1.8%   |
| L3 miss  | 8064 | -    | -    | 4.0%   | 3725 | -    | -    | 7.7%   |
| L2 lines | 227K | -    | -    | 2.9%   | 226K | -    | -    | 5.5%   |
| TLB miss | 9321 | -    | -    | 2.4%   | 9301 | -    | -    | 3.0%   |

## 7.2 Stability of HPC Events

We evaluated HPCs stability in terms of both over-counting and determinism by comparing the data collected from HPCs with data obtained from software instrumentation—results are reported in Table 2. For this experiment, we relied on a basic (single thread) benchmark[12] which computes the first $x$ prime numbers, where $x$ is a user-defined parameter. As a baseline, we used cachegrind [42], which automatically detects the underlying cache structure and builds an equivalent cache model while executing the program. With cachegrind, we can compare the results related to memory accesses and cache misses—in this case, we are also able to assess, to some extent, the accuracy of HPCs. Nevertheless, L3 cache misses, L2 filled lines (it counts opportunistic events at cache line grain and includes prefetcher activity), and TLB miss (we use a specific event that requires the emulation of a second-level TLB) are not available. For these events, we compared the HPCs values of several runs to compute the determinism degree of this source. The results in Table 2 experimentally confirm that, although HPCs could be subject to reliability errors, we have selected events that are more stable and portable across different architectures. Although the L1 miss Err value may be a wake-up call to the reader, it is consistent among the tested architectures and the HPCs variation coefficient. This result stems from cachegrind's inability to model all the hardware counterpart's internal details that vendors do not disclose.

## 7.3 Accuracy of the System-Wide Detection Approach

To assess the capabilities of our detection system, we have performed a system-wide experimental evaluation by building sets of benignware and malware applications. The former relies on the Phoronix Test Suite [34], from which we selected 156 benchmarks (configured with different inputs) showing various behaviors and load profiles. Conversely, to build the set of malicious applications to exercise our solution's capability to detect side-channel attacks, we have not found access to real-world malware of this kind. Consequently, we have crafted such malicious applications starting from the stress-ng suite [32]. We injected side-channel attacks (based on the implementations reported in Table 1) into various benchmarks of the suite, generating a set of 100 malicious applications. The side-channel routine is placed within the benchmark stress function[13]. The attack is anyhow enabled only after a random delay and, after its activation, the side-channel procedure executes with a specific probability—we set this probability to 10%. By introducing these sources of uncertainty, we increased the non-determinism degree that attacks may exploit in realistic scenarios.

---
[12]`sysbench –test=cpu –cpu-max-prime=20000`.
[13]The stress function of each benchmark is called several times into a stress-ng main loop according to input parameters.

As described in Section 5, the behavior of our detection system depends on the $\alpha$, $\beta$, and $\gamma$ hyperparameters. We set $\alpha$ and $\beta$ to 1 for the entire experimental phase while varying $\gamma$ to evaluate the detection according to different threshold levels. As discussed, $\alpha$ and $\beta$ represent the rates that regulate the score progression of each process in the system. By setting $\alpha = \beta = 1$, we are identifying a critical scenario for our detection system, as we slow down the detection of malicious applications while reducing the possibility for a benign application to "recover" from spurious actions being detected as malicious. At the same time, by varying $\gamma$, we somewhat change the responsiveness to an undergoing attack. As stated in Section 4.2, we recall that $\mathcal{S}_1$ and $\mathcal{P}_4$ predicates identify, respectively, side channels directly exploiting the cache levels (L1, L2, LLC) and external caching structures (i.e., TLBs) to manipulate the processor caches indirectly. In our tests, indirect attacks refer to XLATE implementations.

Figure 2 shows the results of the detection accuracy as confusion matrices. The standard benchmarks (i.e., with no side-channel attack injected) are labeled as *OK*, while $\mathcal{S}_1$ and $\mathcal{P}_4$ indicate the direct and the indirect attacks, respectively. Confusion matrices with $\gamma = 1$ illustrate the behavior of our detection system as if the scoring system were not available. In this configuration, any application becomes suspected after a single violation of any metric. As we can observe, the number of false positives is non-minimal, and on the i7-6700HQ, it is even higher than real negatives. Overall, the benchmarks which have been wrongly suspected are the ones that either: i) involve a large number of `forks` and therefore propagate the information associated with the measures across a large number of processes; ii) implement data processing or machine learning algorithms iii) are memory-intensive scientific applications or explicitly test the memory hierarchy.

Nonetheless, the number of false positives quickly decreases as the value of $\gamma$ increases. Indeed, this is related to the fact that subsequent observations can filter out any potential spike in applications' activity without prematurely marking the process as suspected. This trend matches exactly our expectations, also validating the viability of the scoring system. Our experiments did not report any false-negative detection.

The approach we have proposed well fits scenarios in which a higher level of security is desired. However, the system is still prone to performance optimization under very low-security risks. Moreover, by design, the tuning mechanism aims to reduce the likelihood of experiencing false negatives at the cost of slightly increasing the number of false positives. Nevertheless, if $\gamma$ is set to a suitably high value, this number becomes negligible.

By definition, a detection system is not a predictor, but it reacts to some events and makes decisions according to its model. Indeed, such a characteristic is crucial. Before classifying a malicious process as suspected, we expect part of its attack to have been executed—at least, the portion required to generate an identifiable pattern by our detection system. Typical side-channel attacks rely on a preliminary *preparation phase* (e.g., probing the cache) during which no data is actually read. If our detection system can detect a side-channel attack during this preparation phase, the attacker will not be able to read any data. Conversely, if the detection system identifies the attack during its *extraction phase*, then some amount of information might be read by the attacker.

Overall, the amount of data that an attacker can read even if our detection system is active is an important metric to assess the accuracy of our system-wide detection approach. Therefore, we have carried out an experiment to quantify the amount of data that a malicious process can read before its detection. In this experiment, the attacker shares a chunk of read-only memory with the victim and tries to leak information by mounting a side-channel attack on a byte-by-byte basis. Concurrently, the victim reads the shared buffer one byte at a time with some delay among subsequent accesses, generating all the conditions to perpetrate the cache-based attack. In this experiment, we have set the secret's size to 256 bytes—a non-minimal buffer corresponding to the size of a large Advanced Encryption Standard (AES) key—and studied the attack's effectiveness to extract data before being detected. Figure 3 shows the results of this experiment with detection capabilities turned on with different values of $\gamma$ and different victim's read rates[14] . With $\gamma = 100$, our approach can detect the attack before it extracts a significant fraction of the data extracted when no detection was active, but only when the victim reads with small delays. By decreasing $\gamma$, the percentage of correctly extracted data is reduced.

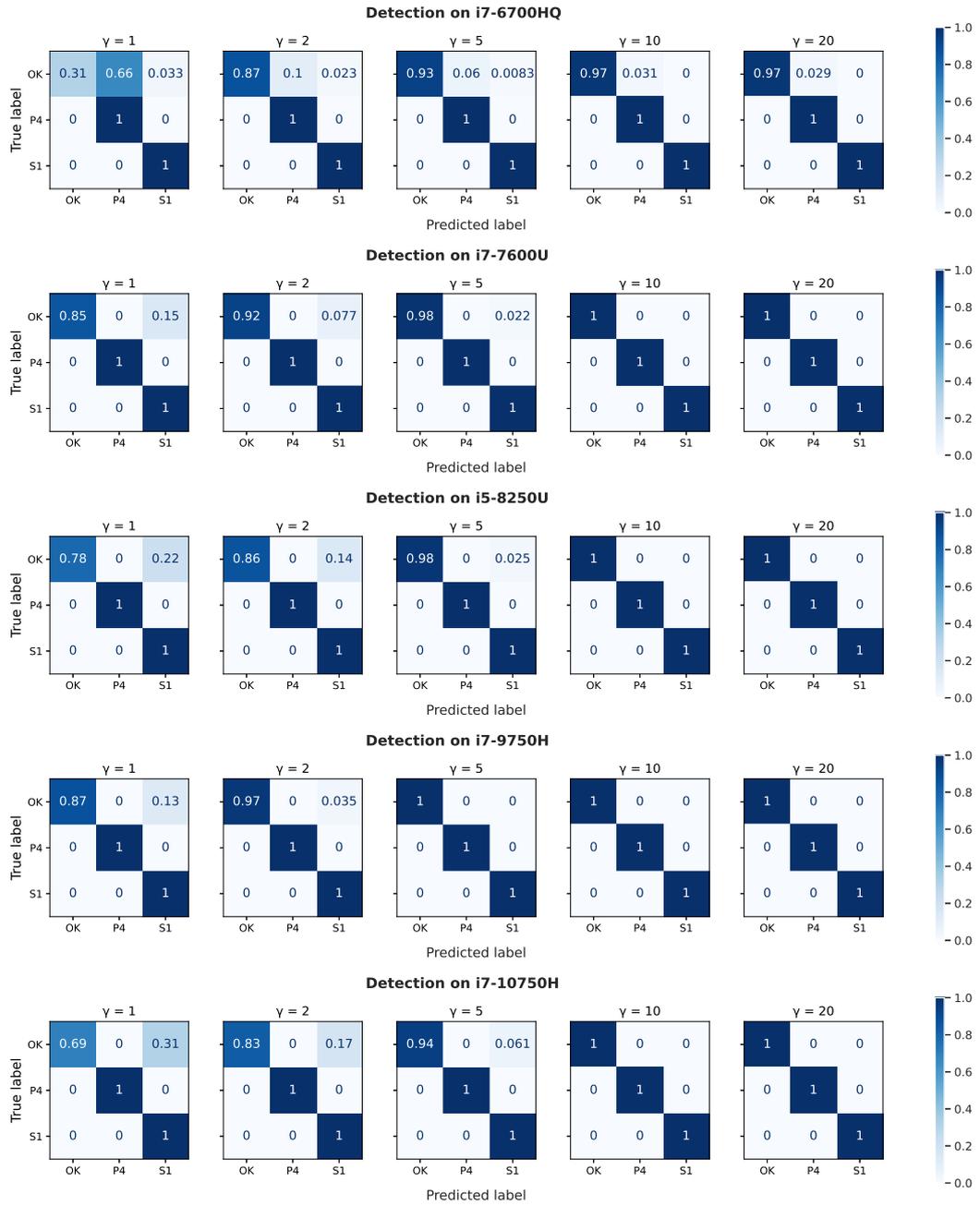

Fig. 2. Detection accuracy evaluation for different values of $\gamma$ ($\alpha, \beta = 1$). $\mathcal{S}_1$ and $\mathcal{P}_4$ indicate the direct and the indirect attacks, respectively, while *OK* indicates normal processes.

Although the reader may think that by increasing the victim's read rate, the detection may fail to identify the attack promptly, our results show that the percentage of extracted bytes decreases for very high-frequency reads on all examined architectures. This phenomenon is due to side-channel attacks being more sensitive to the noise generated when the activity in the system increases.

### 7.4 Performance Assessment

We have studied the performance improvement we can obtain with our monitoring proposal. To quantify the performance benefit of our approach, we have again relied on the Phoronix Test Suite, selecting a set of benchmarks that interact with the system in different ways, according to the following classes of behavior:

(A) intensive disk I/O operations (`compilebench`);
(B) pressure on the scheduler and context switch operation, also considering multithreaded applications (`hackbench`, `ctx_clock`);
(C) a large number of system call invocations, such as `fork`, `exec`, and those related to memory management (`OSBench`);
(D) high usage of the network socket API (`sockperf`);
(E) high usage of the GNU C Library APIs (`glibc-bench`);
(F) complex workloads, related to browsers and databases (`selenium`, `sqlite-speedtest`, `Apache`).

We also note that selecting these benchmarks allows profiling different classes of applications, namely CPU-bound ones (in userspace) or applications that repeatedly interact with the kernel, forcing the application to make a substantial number of mode switches. Given the implementation of our software patches, we should have an influence on the performance of the considered applications. No side-channel attack has been mounted in this experiment.

These benchmarks were run in the four following scenarios to evaluate the performance impact of the system-wide detection scheme, also accounting for the effect of the observation window's length:

(A) Mainline kernel 5.4.145 with KPTI, retpolines, SSB mitigations, and all the patches discussed in Section 6 enabled by default for all processes—referred to as `Generic` in the plots.
(B) Kernel 5.4.145, with our support for dynamic patching, but with system-wide monitoring disabled—referred to as `Monitor OFF` in the plots.
(C) Kernel 5.4.145, with our system-wide detection scheme activated, with an observation window set to $2^{20}$ clock cycles, which was the minimum observation window value considered by the adaptive approach—referred to as `Monitor (short window)` in the plots.
(D) Kernel 5.4.145, with our system-wide detection scheme activated, with an observation window set to $2^{24}$ clock cycles, which was the maximum observation window value considered by the adaptive approach—referred to as `Monitor (long window)` in the plots.

The results for the benchmarks in these configurations are reported in Figure 4, where we show the overhead with respect to the mainline kernel 5.4.145 with no active patch, which is therefore vulnerable to all the discussed attacks—values are averaged over three different runs. By the results, we can observe that the `Monitor OFF` approach offers a performance slowdown with respect to the `Generic` configuration, which is up to 4 orders of magnitude lower while showing an overhead over the unpatched mainline kernel lower than 4% on all architectures and for all application classes. This means that the support we have introduced in the kernel to enable/disable at runtime the various security patches is lightweight and non-intrusive.

---

[14]In this experiment, we have used an observation window of $2^{20}$. A relation between the delay between two victim's reads (in $\mu$sec) and the observation window's size (in clock cycles) can be devised by considering that at a frequency of 1 GHz ($\sim 2^{30}$), we have, given the sampling period of $2^{20}$, $2^{10}$ samples in one second. 1024 samples in one second correspond, roughly, to 1 sample/ms.

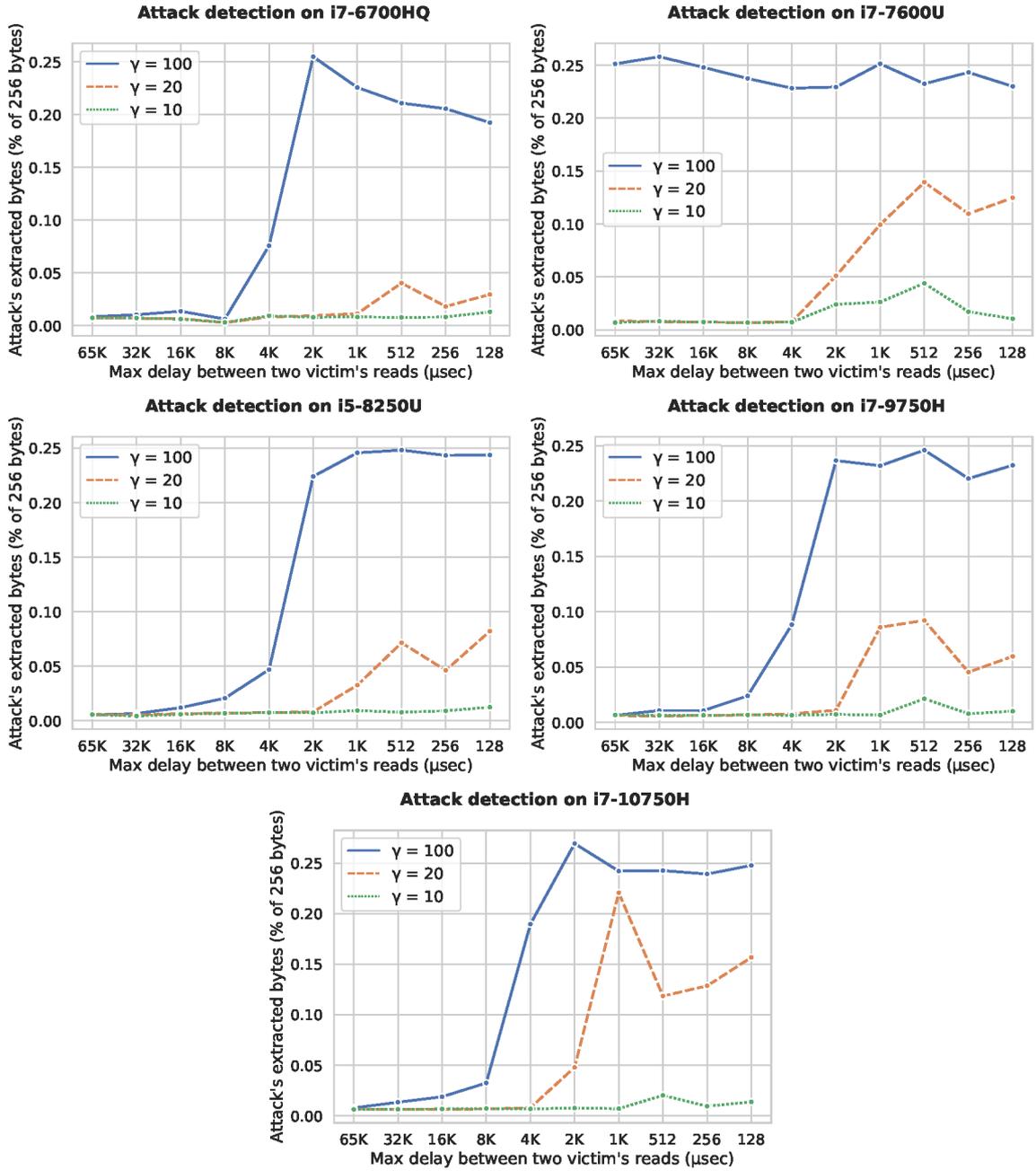

Fig. 3. Percentage of a 256-byte secret that an attack can correctly extract before its detection for different values of $\gamma$ ($\alpha, \beta = 1$) and victim's read rates.

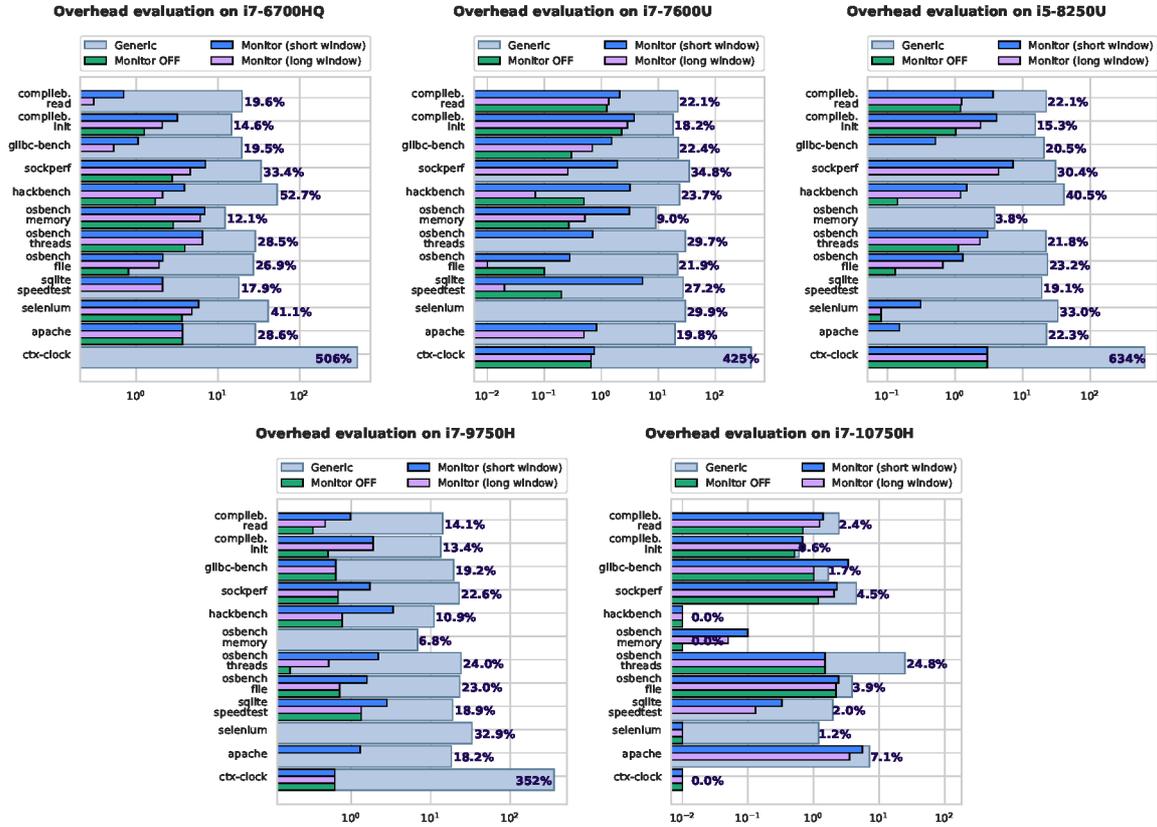

Fig. 4. Performance Effects of the HPC-based Monitoring System on different Architectures (logscale on the $x$ axis).

Conversely, the overhead of the Monitor configuration over the Monitor OFF configuration is negligible.

It is interesting to note that the impact of the window length is minimal: considering that they are related to the maximum/minimum values supported by our system, this experiment shows that the expected overhead, also accounting for the adaptive optimization of the window, is reduced. Of course, this reduced overhead is coupled with our proposal's increased security level. Overall, this is additional evidence of the viability of our proposal.

A similar trend can be observed for all tested architectures (except i7-10750H) and all classes of applications, although with different relative ratios. This indicates the stability of our approach with respect to the performance of applications. The results on the i7-10750H processor do not match the other models' behavior. This is because Intel, starting from the $10^{th}$ generation of its processors, introduced design changes to patch some hardware vulnerabilities. Consequently, the Linux kernel does not require enabling all the software patches (such as KPTI) on these processors with a mitigation of the performance slowdown. Nevertheless, our approach can still detect side-channel attacks on more modern architecture for which a hardware patch has not been proposed, with reduced overhead.

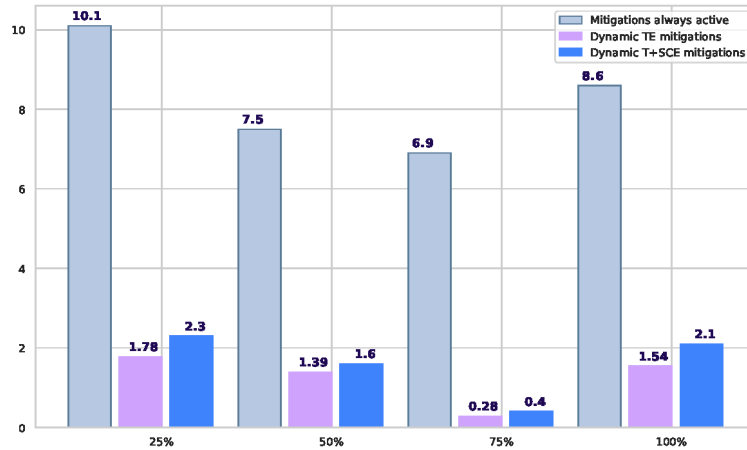

Fig. 5. Performance Penalties by Mitigations on the i5-8250U.

The last experiment we present—the data are reported in Figure 5—relates to an assessment of the overhead due to transient execution mitigations and side-channel mitigations, also when there is significant interference with benignware on the same CPU cores. For this experiment, we only report data taken on the i5-8250U machine for the sake of space. In any case, the results on the other architectures show trends that are perfectly comparable with the data reported on this CPU.

We have launched a number of benchmarks taken from the Phoronix Test Suite equal to the number of available cores on the considered processor. Each benchmark has been statically pinned to one CPU core. We then varied the number of malicious applications, pinned to specific CPU cores, and ran them concurrently with the benignware benchmarks. This setup stress-tests also the per-process detection/mitigation capabilities of our system. We report data associated with the system run with all transient execution mitigations always active (`Mitigations always active` in the plot), with transient execution mitigations activated only for suspected processes (`Dynamic TE mitigations` in the plot), and with transient execution/side-channel mitigation countermeasures activated only for suspected processed (`Dynamic TE+SC mitigations` in the plot). The applications have been selected to avoid any false positive/negative. As in the previous experiment, we report the overhead as the percentage increase over an execution in which no mitigation at all (neither static nor dynamic) is present in the system.

By the result, we observe again that enforcing dynamic mitigations provides a significant overhead reduction, as high as 95%. As expected, the overhead incurred when also SC mitigations are active is higher. Of course, depending on the system's configuration, the user can determine what set of mitigations should be enforced upon the detection of a malware application.

## 8 CONCLUSIONS AND FUTURE WORK

In this paper, we have presented a kernel-level system-wide detection system of cache side-channel attacks based on measures taken from HPCs and relies on decision metrics to deem a (multi-threaded) process as suspected. Our solution has been integrated into the Linux kernel, but it is in principle applicable to other OS kernels. We have coupled our detection system with mitigation actions, both for side-channel attacks and transient execution-based attacks relying on cache side channels to leak data. These mitigations are activated on a fine-grain basis at runtime, as opposed to scenarios where security-oriented tasks are carried out by default independently of the trustworthiness level of the running applications. This allows for better resource usage and improved performance.

The metrics we have devised have allowed us to detect attacks with a negligible percentage of false positives and no false negatives. The data have been collected on different flavors of x86 Intel CPUs and with a comprehensive set of benchmark applications (either benignware or malware).

Based on the comprehensive architecture we have presented in this paper, we plan to expand the set of detection metrics as future work to also account for other kinds of memory-based attacks, such as Rowhammer [31]. Furthermore, we plan to port our solution to processors from vendors other than Intel, e.g. AMD. Concerning the effects of our mitigation strategies, we also plan to carry out an experimental assessment to show the impact on our approach's power consumption.

## ACKNOWLEDGEMENTS

This work is partially supported by the EU Commission in the frame of the Horizon 2020 project SPARTA (Grant #830892).